\documentstyle[prl,epsf,multicol,aps]{revtex}

\begin{document}   
   
\draft   
   
\title{Experimental evidence of delocalized states   
       in random dimer superlattices}  

\author{V.\ Bellani$^1$, E.\ Diez$^2$, R.\ Hey$^3$, L.\ Toni$^4$, 
L.\ Tarricone$^4$, G.\ B.\ Parravicini$^1$, \\ F.\ Dom\'{\i}nguez-Adame$^5$, 
and R.\ G\'omez-Alcal\'a$^6$}
   
\address{$^1$INFM-Dipartimento di Fisica ``A.\ Volta", Universit\'{a}
di Pavia,  I-27100 Pavia, Italy\\   
$^2$GISC, Depto. de Matem\'{a}ticas, Universidad Carlos III   
de Madrid, E-28911 Legan\'es, Spain\\ 
$^3$Paul Drude Institut f\"ur Festk\"orperelektronik,   
Hausvogteiplatz 5-7, D-10117 Berlin, Germany\\  
$^4$INFM-Dipartimento di Fisica, Universit\'{a} di Parma,   
I-43100 Parma, Italy\\   
$^5$GISC, Departamento de F\'{\i}sica de Materiales, Universidad   
Complutense, E-20840 Madrid, Spain\\
$^6$Depto. de Tecnolog\'{\i}as de las Comunicaciones, Universidad   
de Vigo, E-36200 Vigo, Spain}   
   
\date{\today}   
   
\maketitle   
   
\begin{abstract}   
   
We study the electronic properties of GaAs-AlGaAs superlattices with
intentional correlated disorder by means of photoluminescence and vertical
dc resistance.  The results are compared to those obtained in ordered and
uncorrelated disordered superlattices.  We report the first experimental
evidence that spatial correlations inhibit localization of states in
disordered low-dimensional systems, as our previous theoretical
calculations suggested, in contrast to the earlier belief that all
eigenstates are localized.

\end{abstract}   
   
   

\begin{multicols}{2}

\narrowtext

In recent years, a number of tight-binding~\cite{Flores,Dunlap,Science} and
continuous~\cite{Sanchez} models of disordered one-dimensional (1D) systems
have predicted the existence of sets of extended states, in contrast to the
earlier belief that {\em all} the eigenstates are localized in 1D
disordered systems.  These systems are characterized by the key ingredient
that structural disorder is short-range {\em correlated}.  Due to the lack
of experimental confirmations, there are still some controversies as to the
relevance of these results and their implications on physical properties.
In this context, some authors have proposed to find {\em physically
realizable systems\/} that allow for a clear cut validation of the
above-mentioned purely theoretical
prediction~\cite{PRB94a,PRB94b,IEEE,PRB96}.  Given that semiconductor
superlattices (SL's) have been already used successfully to observe
electron localization due to
disorder~\cite{Chomette,Pavesi,Zhang,Lorusso,Mader1,Mader2}, these authors
have suggested SL's as ideal candidates for controllable experiments on
localization or delocalization and related electronic
properties~\cite{PRB94a,PRB94b,IEEE,PRB96}.

To the best of our knowledge, up to now there is {\em no experimental
verification\/} of this theoretical prediction owing to the difficulty in
building nano-scale materials with intentional and short-range correlated
disorder.  However, the confirmation of this phenomenon is important both
from the fundamental point of view and for the possibility to develop new
devices based on these peculiar properties.  In this work we present an
{\em experimental verification\/} of this phenomenon in semiconductor
nano-scale materials, taking advantage of the molecular beam epitaxy growth
technique, which allows the fabrication of semiconductor nanostructures
with monolayer controlled perfection.

We grew several GaAs-Al$_{0.35}$Ga$_{0.65}$As SL's and we studied their
electronic properties by photoluminescence (PL) at low temperature and dc
vertical transport in the dark.  Indeed PL has been proven to be a good
technique to study the electronic properties of disordered
SL's~\cite{Chomette,Pavesi,Zhang}, giving transition energies comparable
with theoretical calculations of the electronic levels.  The electronic
states were calculated using a Kronig-Penney model that has been shown to
hold in this range of well and barrier thicknesses, with precise
results~\cite{Fujiwara}.  This allows the analysis of the experimental
transition energies for PL and the ascertainment of the localization and
delocalization properties of the SL's.  The details of the calculations and
a schematic view of the conduction-band profiles of the three SL's can be
found in Ref.~\onlinecite{IEEE}.


The samples are three undoped SL's grown by molecular beam epitaxy (MBE).
All the SL's have 200 periods and Al$_{0.35}$Ga$_{0.65}$As barriers $3.2\,$
nm thick.  In the {\em Ordered\/}-SL all the 200 wells are identical with
thickness $3.2\,$ nm (hereafter referred to as A wells).  In the {\em
Random\/}-SL, $58$ A wells are replaced by wells of thickness $2.6\,$ nm
(hereafter referred to as B wells) and this replacement is done randomly.
The so-called {\em Random dimer\/}-SL is identical to the {\em Random\/}-SL
with the additional constraint that the B wells appear only in
pairs~\cite{IEEE}.  In the latter sample the disorder exhibits the desired
short-range spatial correlations.  In each sample, the SL is cladded on
each side by $100\,$ nm of n-Al$_{0.3}$Ga$_{0.7}$As, Si doped to $4\times
10^{18}$ cm$^{-3}$, with a $50\,$ nm n-GaAs buffer layer (doped to $4\times
10^{18}$ cm$^{-3}$) on the substrate and a $3\,$ nm n-GaAs cap layer (doped
to $6\times 10^{18}$ cm$^{-3}$).

We measured X-ray diffraction spectra of the SL's with a double-crystal
diffractometer, in order to check their structural parameters.  The
diffraction curve at ($004$) symmetric reflections for the two disordered
samples show satellite peaks of order $\pm 1$ lying close to $\pm 0.8\,$
degrees with respect to the GaAs peak.  These satellite peaks are located
at identical positions for the two disordered SL's, showing that the random
SL's have identical periods.  Therefore, the dimer constraint intentionally
introduced during sample growth is the only difference between {\em
Random\/} and {\em Random dimer\/} samples.

The PL spectra were taken in the $11-300\,$K temperature range with a
closed cycle cryostat, and were excited with $514.5\,$ nm light from an
Ar$^{+}$-ion laser (with an excitation intensity of approximately
$0.5\,$W/cm$^2$).  Photoluminescence was dispersed by a $0.46\,$m Jobin
Yvon monochromator and detected by a cooled photomultiplier using a
standard lock-in technique.

   
Figure~\ref{fig1} shows the PL spectra of the three SL's at $11\,$ K.  We
observed that the energy of the near-band edge peaks depends on the sample,
but the energy shift between them is almost independent of temperature on a
wide range, as shown in Fig.~\ref{fig2}.  The PL peak for the {\em
Ordered\/}-SL, which lies at $1.688\,$ eV, is due to recombination between
electrons in the conduction-band and heavy-holes in the
valence-band~\cite{Exciton}.  We calculated the miniband structure of this
SL with the Kronig-Penney model, using $\Gamma$ effective masses (in units
of free electron mass) $m_{e}^{*}=0.067$ for GaAs and $m_{e}^{*}=0.096$ for
Al$_{0.35}$Ga$_{0.65}$As.  The expected miniband in the conduction-band
lies in the range between $1.68$ and $1.76\,$ eV, measured from the (very
narrow) heavy hole miniband.  This calculation is in good agreement with
the experimental PL spectrum, the calculated lower energy of the miniband
being very close to the energy at which PL intensity rises up.
Figure~\ref{fig3}(a) shows an energy schema of the radiative transitions in
the {\em Ordered\/}-SL.

Let us now analyze the spectrum obtained for the {\em Random\/}-SL.  The PL
peak of this sample shifts towards higher energies compared with the other
two samples.  In the {\em Random\/}-SL the intentional disorder introduced
by the random distribution of wells B ($2.6\,$ nm) localizes the electronic
states~\cite{IEEE}.  The calculated energy for the transition between
electron and holes in this case is $1.72\,$ eV assuming that the exciton
binding energy is the same in the three SL's.  This value is again in
excellent agreement with the PL peak, as can be seen in Fig.~\ref{fig1}.
Figure~\ref{fig3}(b) shows a schematic diagram of the radiative transitions
between localized states in the {\em Random\/}-SL.

The PL peak of the {\em Random dimer\/}-SL is at $1.696\,$ eV and, as can
be clearly seen in Fig.~\ref{fig1}, is {\em red shifted\/} with respect to
the PL peak for the {\em Random\/}-SL.  As has been shown by
Fujiwara~\cite{Fujiwara}, the {\em red shift\/} of the PL peak in
semiconductor SL's is due to the formation of a miniband with tunnel
process for carriers between the GaAs wells.  This result strongly supports
previous theoretical predictions of the occurrence of a band of extended
states in {\em Random dimer\/}-SL's~\cite{IEEE}.  We calculated the
transmission coefficient for the {\em Random dimer\/}-SL according to
Ref.~\onlinecite{IEEE} and found that the energy difference between the
onset for electron delocalization (that is, the energy at which
transmission suddenly rises, as can be seen in Fig.~\ref{fig4}) and the
heavy hole miniband is around $1.70\,$ eV, in good agreement with the
experimental PL peak energy.  Figure~\ref{fig3}(c) presents a schematic
diagram of the radiative transitions in the {\em Random dimer\/}-SL.

Additionally, the PL line-width gives support to these findings.  The PL
full width at half maximun of the {\em Ordered\/}-SL is $9.1\,$ meV,
increasing to $13.2\,$ meV and $12\,$ meV in the {\em Random\/}-SL and {\em
Random dimer\/}-SL respectively, indicating that these last two samples
reflect in the optical spectra their intentional disorder.

To confirm the above interpretation of the PL spectra we have performed
additional measurements of the resistance at low temperatures~\cite{Koch}.
The results for the temperature dependence of the resistance are shown in
Fig.~\ref{fig5}.  The resistance of the {\em Random dimer\/}-SL is very
similar to the resistance of the {\em Ordered\/}-SL for any temperature
below $40\,$K, and the small differences are due to the different
miniband-width between both SL's (see Fig.~\ref{fig3}).  On the other hand,
the {\em Random\/}-SL shows a much higher resistance in this range of
temperatures.  This is completely consistent with the above interpretation
of the PL spectra and it is clear evidence of the presence of extended
states in the {\em Random dimer\/}-SL showing transport properties very
similar to an {\em Ordered\/}-SL.  Moreover, the resistance of the {\em
Random\/}-SL still depends on temperature below $30\,$K, while the
resistance of the two other samples reaches a plateau.  For low
temperatures, transport properties in the presence of true extended states
should be independent of temperature~\cite{Koch} and, as can be seen in
Fig.~\ref{fig5}, this behavior is only observed in the {\em Random
dimer\/}-SL and in the {\em Ordered\/}-SL, which is additional evidence of
the presence of extended states in these samples.

   
In summary, we have observed that the introduction of short-range
correlations in a disordered semiconductor SL inhibits localization and
gives rise to extended states, as expected theoretically~\cite{IEEE}.  The
positions of the electronic levels were calculated with the Kronig-Penney
model and the calculations show that the {\em Ordered\/}-SL and the {\em
Random dimer\/}-SL exhibit extended electronic states.  According to
theoretical studies, these extended states in {\em Random dimer\/}-SL's are
not Bloch-like, as occurs in {\em Ordered\/}-SL's.  The PL of the {\em
Random dimer\/}-SL is {\em red shifted\/} with respect to the PL of the
{\em Random\/}-SL, indicating the formation of delocalized extended states.
The experimental PL energies are in very good agreement with the calculated
electronic states.  The temperature dependence of the resistance of the
{\em Random dimer\/}-SL is very similar to that of the {\em Ordered\/}-SL.
Both SL's shows no temperature dependence below $30\,$ K as should be
expected for transport in the presence of extended states.  On the
contrary, the resistance of the {\em Random\/}-SL is much higher for any
temperature and shows temperature dependence as would be expected for
localized states.  To conclude, we have experimentally validated the
existence of extended states in low-dimensional random systems with
short-range correlations, where Anderson localization is inhibited.

\acknowledgments   
   
Work in Italy has been supported by the INFM Network ``Fisica e Tecnologia
dei Semiconduttori III-V'' and in Madrid by the CAM under Project
No.~07N/0034/1998.  E.~D.\ and F.~D.-A.\ thank A.\ S\'{a}nchez for
collaboration on these topics during these years.

\end{multicols}
     
\begin{figure}   
\caption{PL spectra of the {\em Ordered\/}-, {\em Random\/}- and   
{\em Random dimer\/}-SL at $11\,$ K.}   
\label{fig1}   
\end{figure}   
  
\begin{figure}   
\caption{Temperature dependence of the PL peak position of the 
{\em Ordered\/}-, {\em Random\/}-  and {\em Random dimer\/}-SL.}   
\label{fig2}   
\end{figure}   

\begin{figure}   
\caption{Schematic diagram of radiative transitions in a) {\em Ordered\/}-SL,  
b) {\em Random\/}-SL and c) {\em Random dimer\/}-SL at $11\,$ K.}   
\label{fig3}   
\end{figure}   
  
\begin{figure}   
\caption{Calculated transmission coefficient as a function of the electron 
energy in the {\em Random dimer\/}-SL. Energy is measured from the heavy hole  
miniband.}   
\label{fig4}   
\end{figure}   

\begin{figure}
\caption{Temperature dependence of the dc resistance of the
{\em Ordered\/}-, {\em Random\/}-  and {\em Random dimer\/}-SL.}
\label{fig5}
\end{figure}
 

\begin{references}
\bibitem{Exciton} Exciton binding energies are  much smaller than the 
        observed peak shifts.
\end{references}
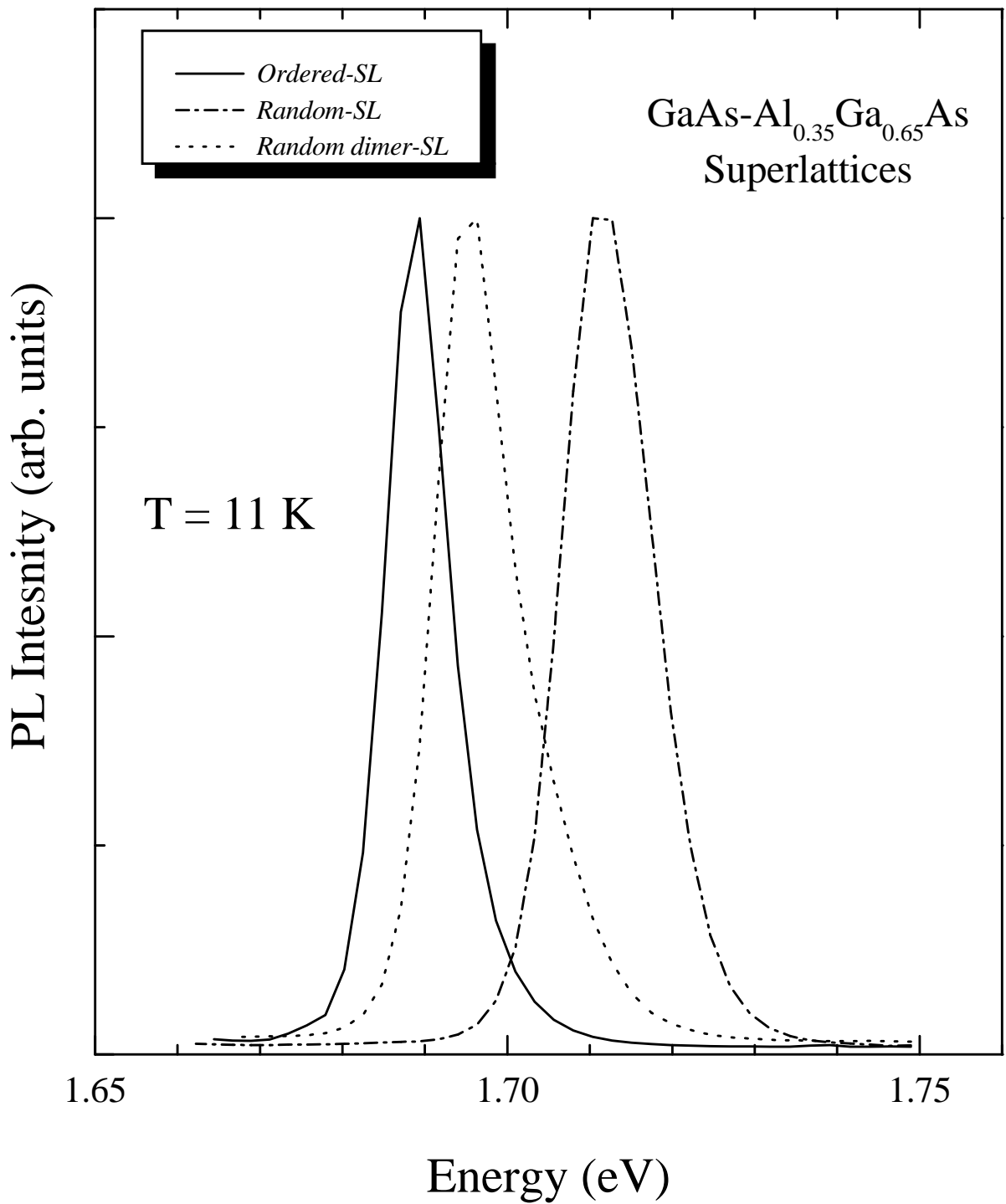
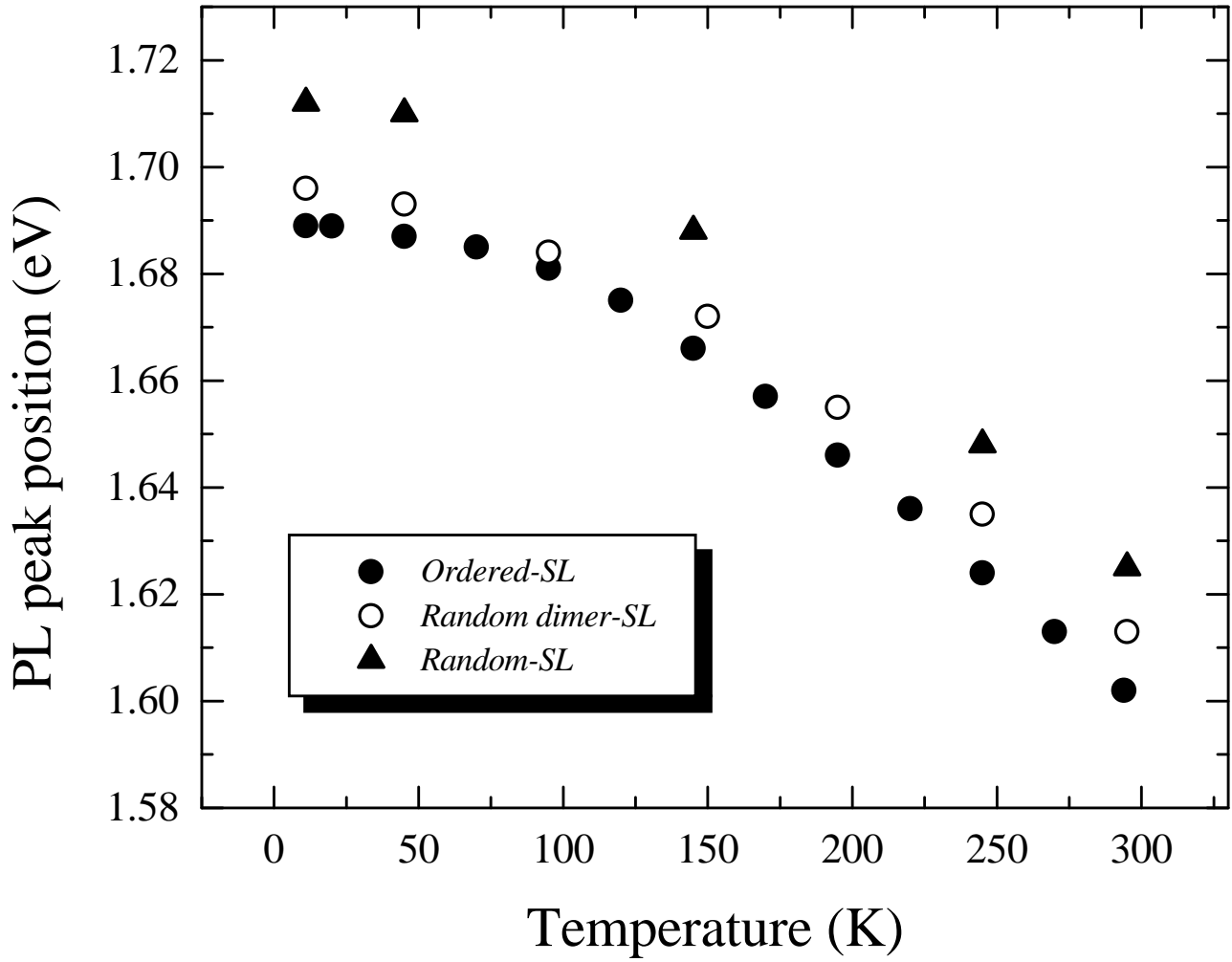
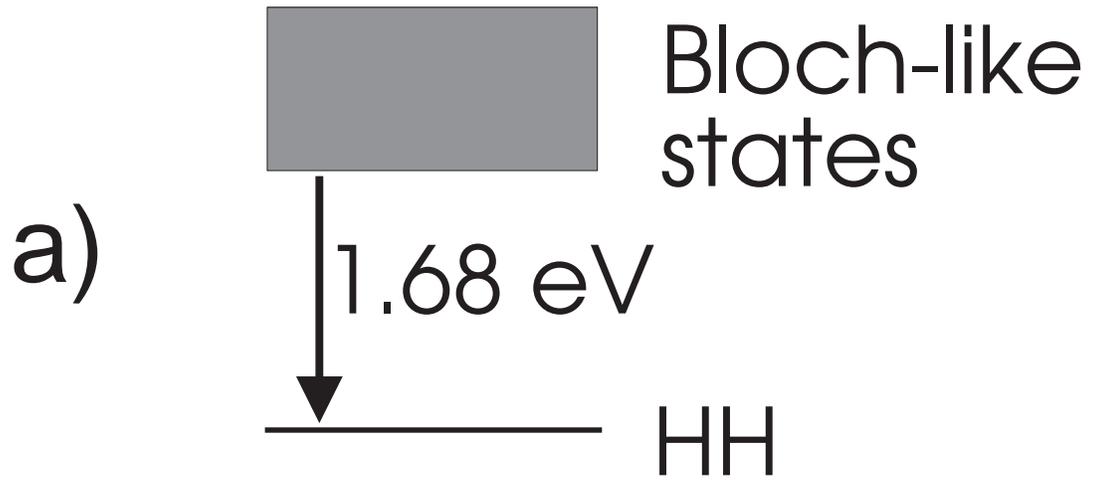
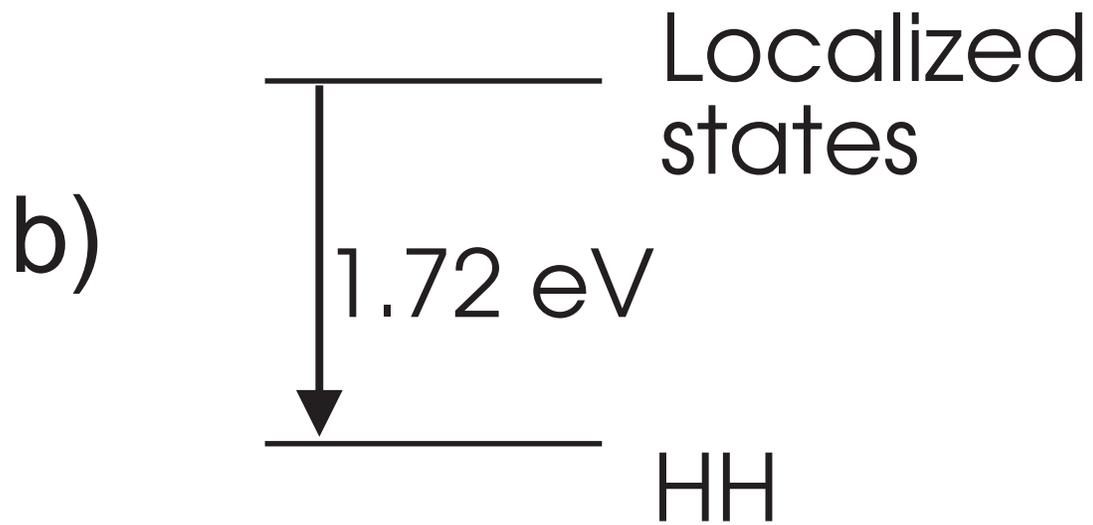
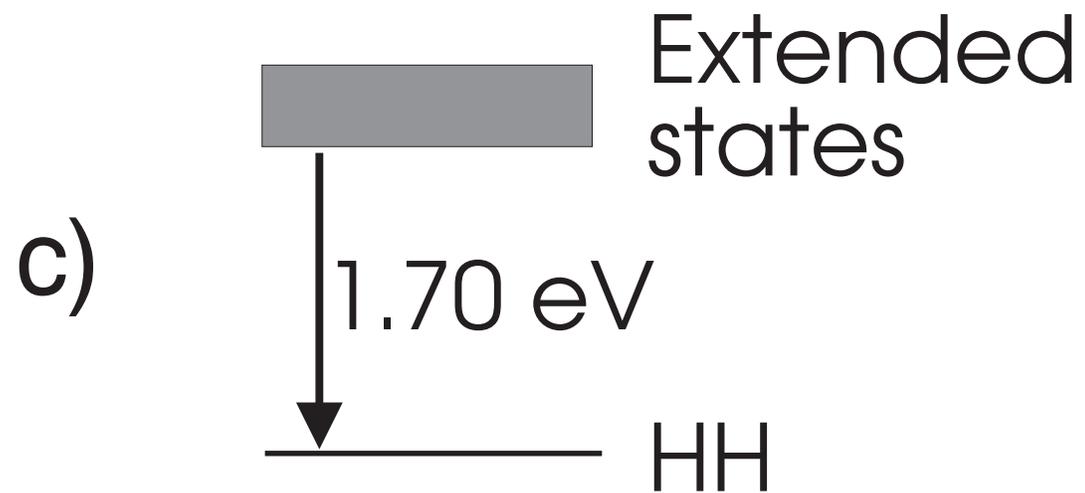
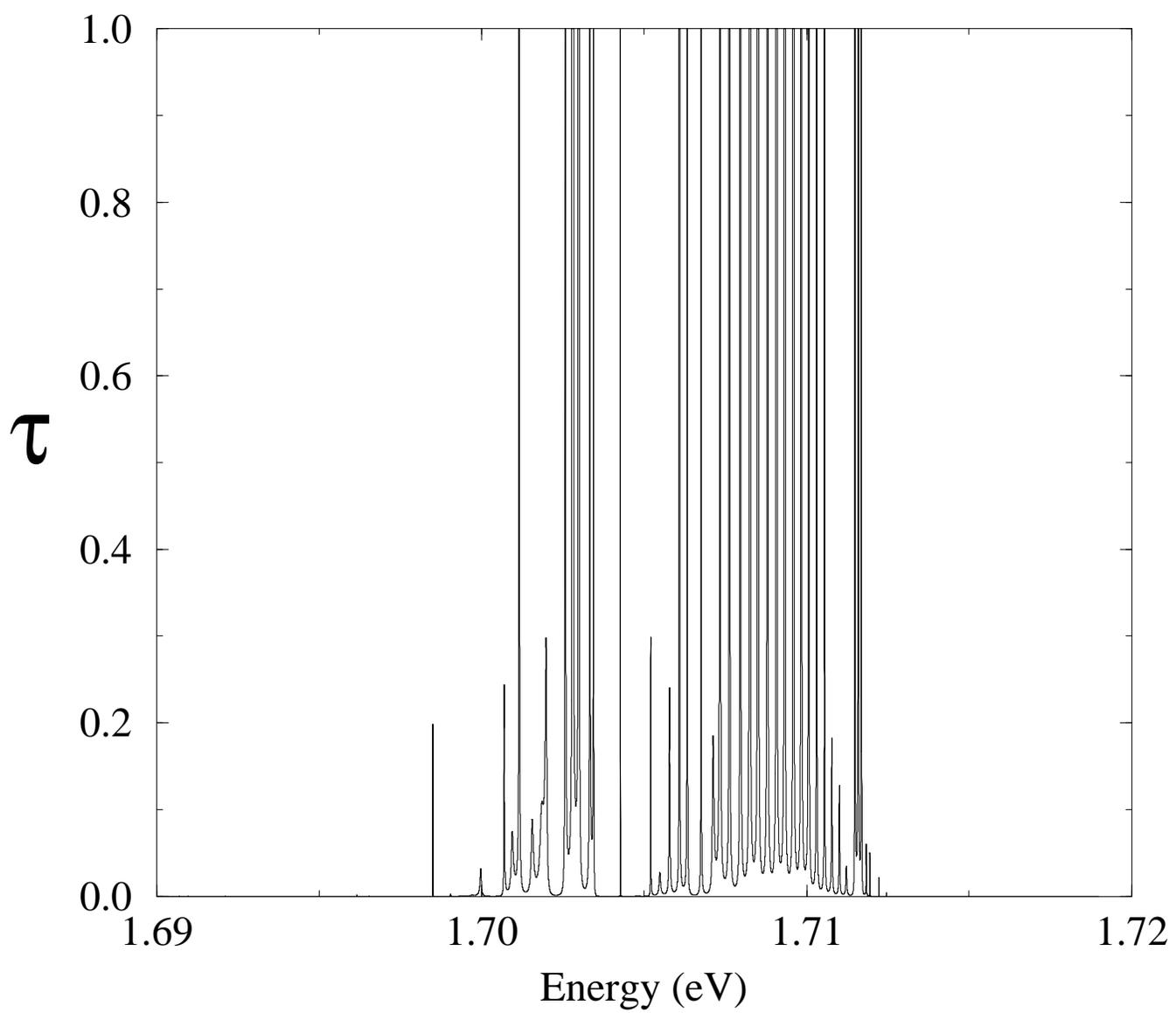
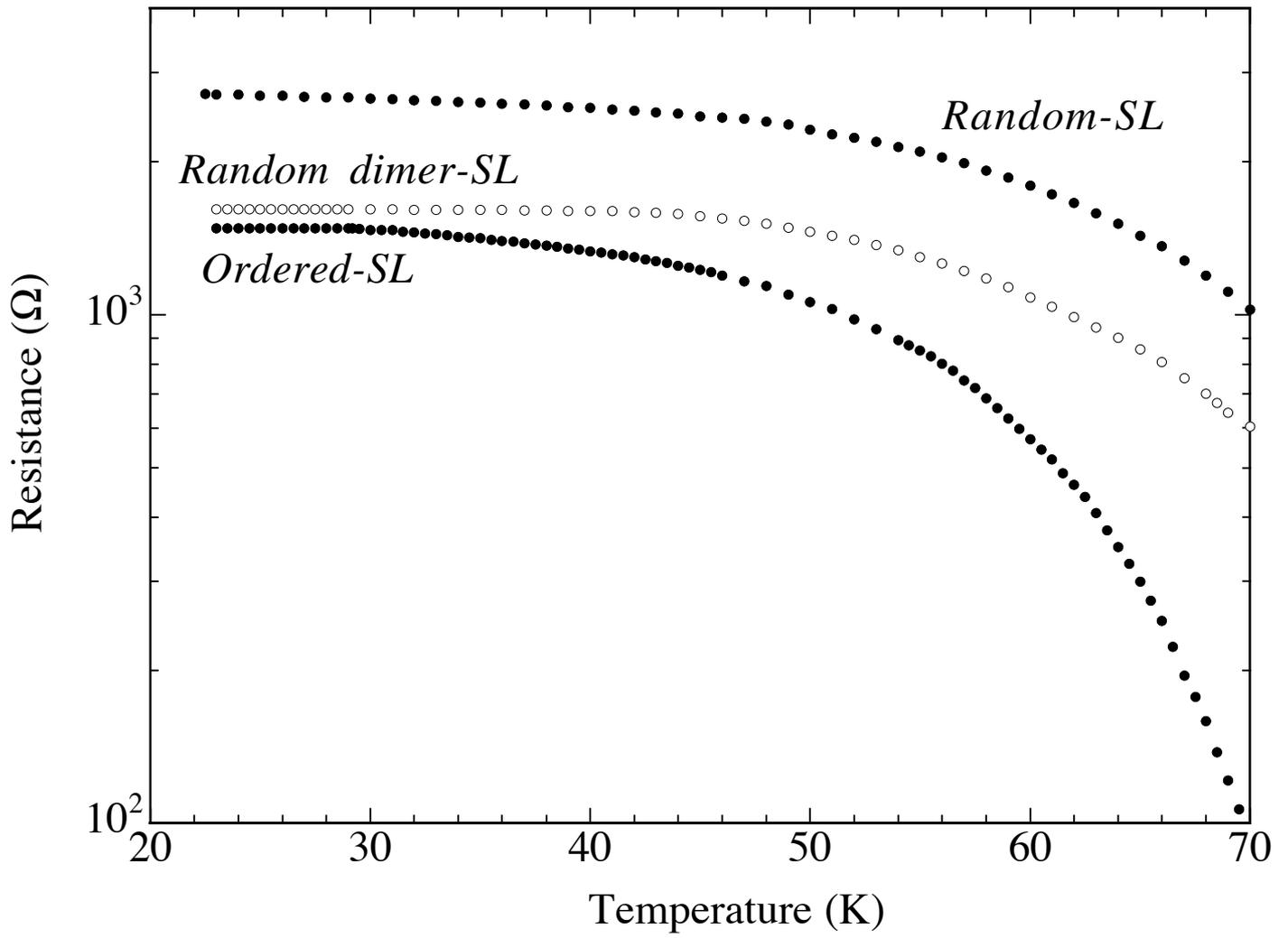
\end{document}